# Astro2020 Science White Paper

## Spectroscopic Observations of the Fermi Bubbles

**Thematic Areas:** ☐ Planetary Systems  ☐ Star and Planet Formation
☐ Formation and Evolution of Compact Objects  ☐ Cosmology and Fundamental Physics
☐ Stars and Stellar Evolution  ☐ Resolved Stellar Populations and their Environments
☒ Galaxy Evolution  ☐ Multi-Messenger Astronomy and Astrophysics

**Principal Author:**
**Andrew J Fox**, Space Telescope Science Institute, afox@stsci.edu, 410 338 5083

**Co-authors:**
**Trisha Ashley**, Space Telescope Science Institute, tashley@stsci.edu
**Robert A. Benjamin**, University of Wisconsin-Whitewater, benjamir@uww.edu
**Joss Bland-Hawthorn,** University of Sydney, jbh@phys.usyd.edu.au
**Rongmon Bordoloi,** North Carolina State University, rbordol@ncsu.edu
**Sara Cazzoli,** Institute of Astrophysics of Andalucia, Spain, sara@iaa.es
**Svea S. Hernandez**, Space Telescope Science Institute, sveash@stsci.edu
**Tanveer Karim,** Harvard University, tanveer.karim@cfa.harvard.edu
**Edward B. Jenkins,** Princeton University Observatory, ebj@astro.princeton.edu
**Felix J. Lockman**, Green Bank Observatory, jlockman@nrao.edu
**Tae-Sun Kim,** University of Wisconsin-Madison, kim@astro.wisc.edu
**Bart P. Wakker,** University of Wisconsin-Madison, wakker@astro.wisc.edu

**Abstract**:
Two giant plasma lobes, known as the Fermi Bubbles, extend 10 kpc above and below the Galactic Center. Since their discovery in X-rays in 2003 (and in gamma-rays in 2010), the Bubbles have been recognized as a new morphological feature of our Galaxy and a striking example of energetic feedback from the nuclear region. They remain the subject of intense research and their origin via AGN activity or nuclear star formation is still debated. While *imaging* at gamma-ray, X-ray, microwave, and radio wavelengths has revealed their morphology and energetics, *spectroscopy* at radio and UV wavelengths has recently been used to study the kinematics and chemical abundances of outflowing gas clouds embedded in the Bubbles (the nuclear wind). Here we identify the scientific themes that have emerged from the spectroscopic studies, determine key open questions, and describe further observations needed in the next ten years to characterize the basic physical conditions in the nuclear wind and its impact on the rest of the Galaxy. Nuclear winds are ubiquitous in galaxies, and the Galactic Center represents the best opportunity to study the constitution and structure of a nuclear wind in close detail.

## 1) Introduction

The Galactic Center (GC) is home to the closest supermassive black hole, Sgr A*, and a surrounding region of intense nuclear star formation, containing many young massive clusters (e.g. Melia & Falcke 2011). Together, these twin energy sources power the nuclear feedback that drives matter out into the halo of the Galaxy. The evidence for this feedback is provided by the spectacular Fermi Bubbles, two giant plasma lobes extending 10 kpc into both Galactic hemispheres and emitting radiation across the electromagnetic spectrum (see Figure 1). This emission ranges from gamma rays (Su+ 2010, Dobler+ 2010, Crocker & Aharonian 2011, Ackermann+ 2014), X-rays and mid-IR (Bland-Hawthorn & Cohen 2003), microwave (Finkbeiner 2004), to radio waves (Sofue & Handa 1984; Carretti+ 2013). In an AGN-driven scenario, the Bubbles are a few Myr old, matching the kinematic age of the nuclear wind (Fox+ 2015, Bordoloi+ 2017) supporting Sgr A* as the power source. The Fermi Bubbles provide a local example of nuclear feedback from a large spiral galaxy and an opportunity to study its effects on the galactic environment. Nuclear winds are ubiquitous in galaxies, and we can derive much more detail on the constitution and structure of the Bubbles than we will ever be able to derive on winds in even nearby galaxies. In this white paper, we identify the scientific themes that have emerged from *spectroscopic* studies of the Fermi Bubbles, and outline directions and key questions for future research. As we show below, our proposed UV and radio spectroscopic investigations are complementary to *imaging* studies in other wavelengths.

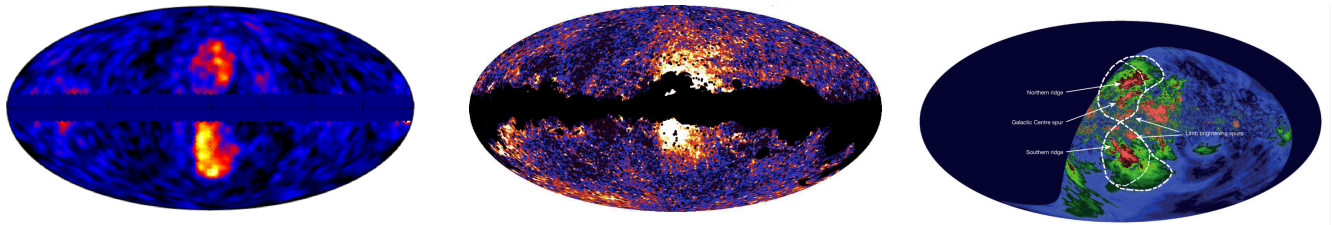

Figure 1: **Left**: the Fermi Bubbles, in an all-sky gamma-ray intensity map (Su+ 2010). The Galactic disk has been masked out and the Bubbles are visible in orange. **Middle**: the "WMAP haze" (Finkbeiner 2004), showing the microwave counterpart to the Bubbles. **Right**: polarized radio emission (Carretti+ 2013).

## 2) Existing Spectroscopic Observations of the Fermi Bubbles

A deficiency of H I exists in the inner Milky Way, particularly within 2 kpc of the Galactic Center, where a clear cavity is seen in 21 cm maps (see Figure 2, left; Lockman 1984, Lockman & McClure-Griffiths 2016). Within this cavity, sensitive observations with the Green Bank Telescope made in the last decade have revealed a population of around 200 compact H I clouds (Figure 2, right; McClure-Griffiths+ 2013; di Teodoro+ 2018). These clouds have latitudes within ±10° latitude of the Galactic plane, temperatures of ~8000 K, typical H I column densities of ~$10^{19}$ cm$^{-2}$, and kinematics that suggest they are being entrained in a biconical nuclear wind. They appear to represent neutral clumps swept up by the hot outflowing wind into the Fermi Bubbles. We still have a limited understanding of their origin, fate, and existence outside the regions studied.

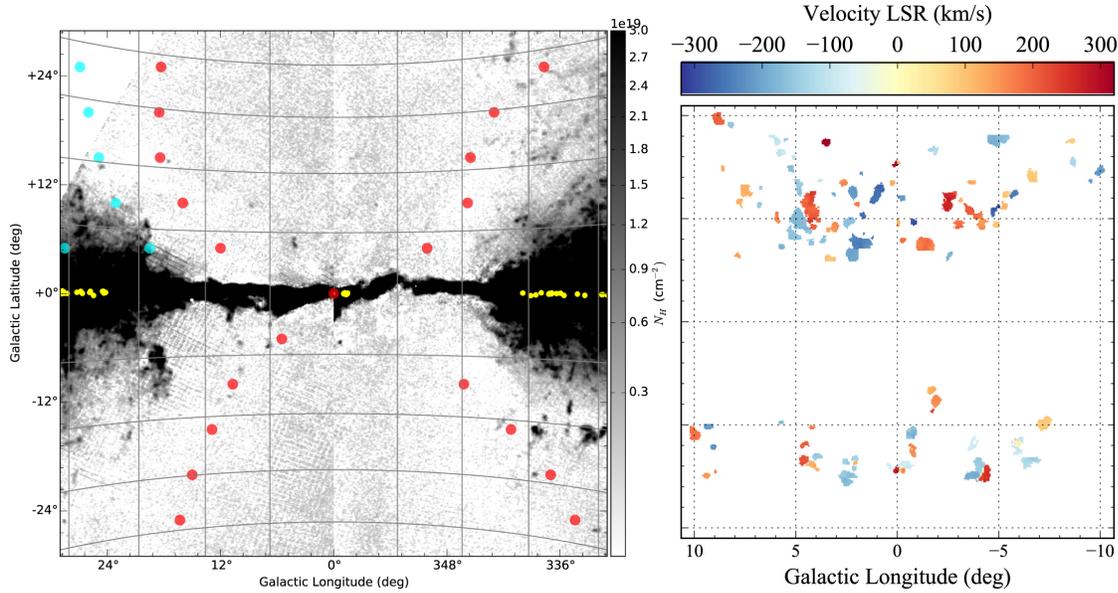

Figure 2: **The radio view of the Galactic Center. Left**: 21cm H I column density map of the Galactic Center showing the absence of H I in the inner Galaxy (Lockman & McClure-Griffiths 2016). The grid marks 1 kpc intervals in distance from the Galactic center and distance from the plane. The red dots show the outline of the Fermi Bubbles. **Right**: Deeper map of 21 cm clouds in the inner 20x20 degree region of the Galaxy (di Teodoro+ 2018), color-coded by LSR velocity, revealing a population of compact H I clouds. These clouds are thought to be swept-up in the nuclear wind.

UV absorption-line observations can be used to probe the ionized component of the nuclear wind. The UV offers many diagnostics of the physical and chemical conditions in the wind, over a wide range of ionization states. This includes tracers of warm ionized gas at $T$~$10^4$ K (O I, N I, C II, Si II, Si III, S II, and Fe II) and highly-ionized gas at $T$~$10^5$ K (C IV, Si IV, and N V). By targeting background AGN lying behind the Fermi Bubbles with the Cosmic Origins Spectrograph (COS) and Space Telescope Imaging Spectrograph (STIS) on the *Hubble Space Telescope*, many authors have studied the properties of embedded gas clouds in UV absorption (Keeney+ 2006, Zech+ 2008, Fox+ 2015, Bordoloi+ 2017, Karim+ 2018). Comparisons of the covering fraction of high-velocity absorption inside and outside the Bubble have shown an enhanced incidence of absorption inside the Fermi Bubble (Bordoloi+ 2017, Karim+ 2018), though for individual directions cloud distances are generally unknown. These studies have led to knowledge of the spatial extent, kinematics, chemical abundances, and physical conditions of the nuclear wind.

A promising and complementary avenue of research for studying the Fermi Bubbles in absorption is the use of *stellar targets* near (or beyond) the Galactic Center, rather than AGN. Massive (OB) and blue horizontal branch (BHB) stars are suitable types. If the stars have good distance constraints, either from *Gaia* or spectral-typing, then the distance to absorbing clouds seen in the stellar spectra can be bracketed. Furthermore, if *pairs* of stars (or star-AGN pairs) at small angular separation are identified, then the comparison of the UV absorption in the two directions can provide key distance information on the absorbing gas (see Figure 3, Savage+ 2017, showing the only published example of a foreground-background pair near the Galactic Center). Finally, stellar sources have the advantage of no IGM contamination, which affects AGN spectra (particularly for high redshift sources), and thus using stars simplifies the absorption-line analysis.

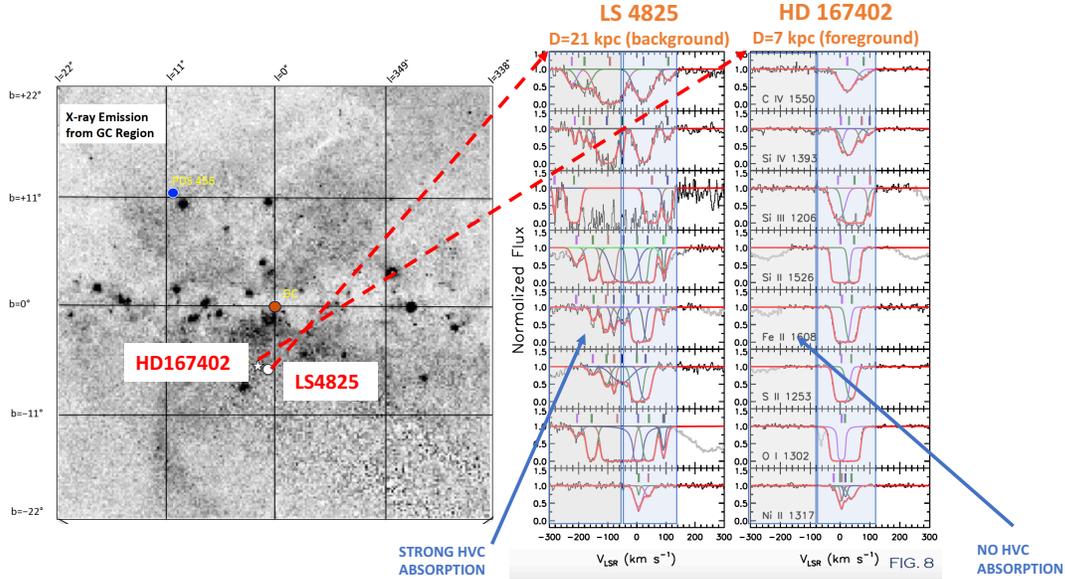

Figure 3: *HST*/STIS ultraviolet spectra of a pair of supergiant stars near the Galactic Center. One star (LS 4825) lies in the background at a distance d=21 kpc; the other (HD 167402) lies in the foreground at d=7 kpc. The left panel shows a ROSAT X-ray image of the Galactic Center region, with the location of each direction. A comparison of the spectra reveals additional absorption in many UV metal lines in the velocity range –260 to –60 km s$^{-1}$ in the background-star spectrum (shaded gray), tracing gas that lies unambiguously beyond 7 kpc. Adapted from Savage+ (2017).

### 3) Limitations of Existing Spectroscopic Studies of the Fermi Bubbles

Current UV studies of the Galactic Center environment are limited by a shortage of UV background sources bright enough to be observable with existing instrumentation. Of particular note is the lack of low-latitude AGN (|*b*|<30°), which can be used as backlights for studying the wind at small impact parameter from the Galactic disk. Such AGN are heavily extinguished by foreground dust. This situation could be improved with future instrumentation with higher UV sensitivity and with multiplexing capability. The proposed LUMOS spectrograph on *LUVOIR* would open up the study of *multiple* background sources (massive stars and AGN) in the GC region in each pointing. With a higher density of sources covered (LUMOS can easily reach FUV magnitudes of 20 with good S/N and spectral resolution, as opposed to ~18 with *HST*/COS/G130M), one could **map out the morphology, kinematics, and chemical abundances in the foreground absorbing gas more efficiently and comprehensively than is possible with *HST***. We illustrate this possibility in Figure 4, which shows a 10-fold increase in the number of AGN reachable with LUVOIR (though fainter sources tend to be at higher redshift, with more confusion from IGM lines).

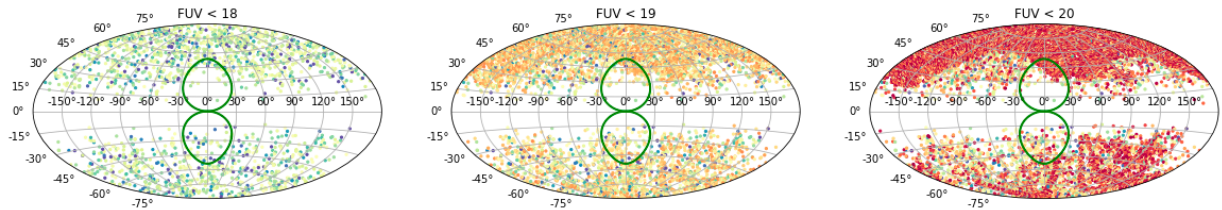

Figure 4: Illustration of the number of QSOs as a function of GALEX FUV magnitude on an all-sky map in Galactic coordinates, with the Fermi Bubbles outlined in green. The AGN are drawn from the Million quasar catalog (v5.7; Flesch+ 2015). Reaching FUV=20 provides a 10-fold increase in the density of sources compared to FUV=18. This enables many more Fermi Bubble directions to be observed, particularly at low latitudes in the south.

## 4) Key Properties to be Constrained with Future Observations

In this section, we identify several key properties of the Galactic Center wind, each of which is poorly constrained. We discuss future observations and simulations needed to make progress.

1. **Chemical Composition and Dust**. What is the metallicity and dust content of the outflowing clouds in the nuclear wind, and what does this tell us about cloud origin and cloud survival? Clouds that are swept up by the wind should contain dust; those that condense out of it should not. Dust can be measured using elemental depletion measurements from *UV absorption-line ratios* (e.g. Cr II/Zn II, Fe II/O I) *from current and future missions* (*HST, LUVOIR, HabEx*) which compare the column densities of depleted to non-depleted elements. Dust not only hides mass, but in a radiation-driven outflow, dust couples electrostatically with radiation to push the swept-up material out into the halo.

2. **Morphology**. What are the morphologies and sizes of the outflowing clouds in the nuclear wind? Do they show head-tail structures indicating an interaction with a surrounding medium, and do they differ from other H I clouds near the Galactic Center (Winkel+ 2011)? Is there an evaporation effect in which the cloud mass decreases outward, as in the winds of nearby star-forming galaxies (Pereira-Santaella+ 2016)? High-resolution hydrodynamic simulations are now modeling cool gas in AGN-driven and star-formation driven outflows (Costa+ 2015, Banda-Barragan+ 2019, Schneider+ 2018; **figure on right, showing a 25 Myr old galaxy outflow**). *Deep 21 cm mapping* with new and upgraded radio facilities is needed to determine the observed properties of the cool clouds and constrain the physics included in the simulations. 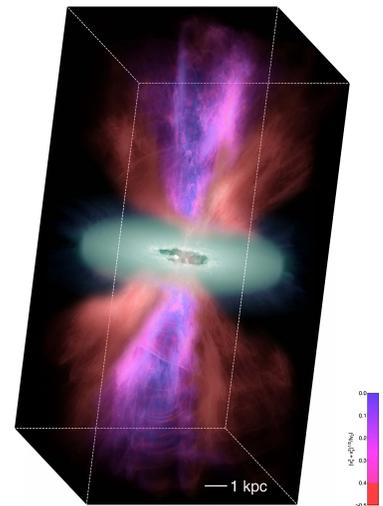

3. **Kinematics**. Does the nuclear wind show a decelerating or accelerating velocity profile? The MW is the only galaxy where the wind profiles can be observed. Currently, H I studies show tentative evidence for acceleration at $|b|<10°$ (Di Teodoro+ 2018), whereas UV studies appear to show deceleration at $|b|=20–50°$ (Bordoloi+ 2017, Karim+ 2018). This may be because the H I and UV observations trace different thermal phases of the wind, or because they trace different latitudes. *Further UV (HST, LUVOIR) and radio (GBT, Parkes) observations* at intermediate latitudes would help resolve this discrepancy.

4. **Age of the Fermi Bubbles**. Ages derived from UV kinematics provide independent and orthogonal measurements compared to imaging studies. The age is key in determining whether the FB emission is leptonic or hadronic (Crocker+ 2015, Mou+ 2018). UV kinematic age estimates are ~6–9 Myr (Fox+ 2015, Bordoloi+ 2017), supporting an AGN-driven model for the Fermi Bubbles, but these studies are based on small numbers of sightlines. Further refinements to these ages from *expanded UV samples* are needed.

5. **Mass outflow rate**. What is the mass outflow rate in the nuclear wind? How does it compare to the mass outflow rate from the rest of the galaxy? This will reveal whether the Milky Way expels more gas in the nuclear region or across the disk (nuclear feedback vs distributed feedback). Radio and UV observations are both needed, since the mass flow rates depend on summing over the neutral and ionized phases of the wind. Existing estimates for the mass outflow rate are ~0.1 $M_\odot$ yr$^{-1}$ in the neutral gas (di Teodoro+ 2018) and ~0.2–0.4 $M_\odot$ yr$^{-1}$ in the warm (~$10^4$ K) component (Bordoloi+ 2017). These numbers need to be calculated with larger samples and including other thermal phases.

6. **Molecular gas**. Although we already have a picture of the neutral and ionized phases of the wind (via the radio and UV observations, respectively), we are missing any information on the molecular phase of the Galactic Center wind. Pathways for detecting molecules include searching for OH emission in the 18 cm radio lines, and looking for near-IR emission from the H2 1–0 S(0) and H2 1–0 S(1) lines, at 2.224 and 2.122 microns. These lines are often observed in the nuclear winds from other galaxies (e.g. Emonts+ 2014, Pereira-Santaella+ 2016). They are observable from the *ground with integral-field units (e.g. Gemini/NIFS)*, and will be available with *space-based observations on JWST*.

7. **Warm ionized gas**. What can we learn about the Galactic Center wind from H$\alpha$ observations? For a photoionized cloud, the H$\alpha$ intensity provides a direct measure of the ionizing radiation field. Shocks may also be important in exciting H$\alpha$ in the energetic Galactic Center environment. Observations from *optical spectroscopic facilities*, such as the Wisconsin H-alpha Mapper (WHAM), are needed to study H$\alpha$ emission from the Galactic Center, to identify the ionization mechanism, and to constrain the mass of warm ionized gas. Low-extinction windows allow H$\alpha$ to be studied close to the Galactic Center.

## 5) Broader Impact and Concluding Remarks

The Fermi Bubbles are our closest example of AGN feedback. They provide a golden opportunity for detailed studies of the recycling of matter between the nucleus of a spiral galaxy and its halo, and the close proximity of the Galactic Center enables spatially-resolved observations across the electromagnetic spectrum. Such observations are not possible in such detail in any other galaxy. Multi-wavelength observations have been (and will continue to be) essential to build a comprehensive observational view of the Fermi Bubbles. In this white paper, we have emphasized the importance of the UV and radio spectral ranges, each of which contains crucial diagnostic spectral lines, which are essential to understand the kinematics, chemical abundances, and physical conditions in the wind. We have also highlighted the potential use of radio and far-IR emission lines for tracing any molecular gas in the wind, which is currently a key missing phase. Insights gained by studying the multi-phase Galactic wind, where we can resolve detail and spatial variation, will be applicable to the study of winds in external galaxies. Finally, various groups have attempted to claim new physics, e.g. dark matter decay signals, from the Galactic Center region. We need to understand the basic high-energy astrophysical processes relating to the GC before new physics can be claimed. The observations described in this white paper will establish a robust set of observations on the Fermi Bubbles and the broader Galactic Center region, and will therefore provide such an empirical foundation.